\documentstyle[12pt]{article}

\begin{document}
\begin{titlepage}

\hfill   \hbox{\bf SB/F/95-233}

\hfill     \hbox{\bf  hep-th/9603013}

\hrule
\vskip 1.5cm
\centerline{\bf NON-ABELIAN BF THEORIES WITH  SOURCES }
\centerline{\bf  AND 2-D GRAVITY} 
\vskip 1cm

\centerline{J. P. Lupi$^1$, A. Restuccia$^1$ and J. Stephany$^{1,2}$}
\vskip 4mm
\begin{description}
\item[]{\it $^1$ Universidad Sim\'on
Bol\'{\i}var,Departamento de F\'{\i}sica,Apartado Postal
89000,  Caracas 1080-A, Venezuela.} 
\item[]{\it  $^2$ Centro Brasilero de Pesquisas F\'{\i}sicas,
Departamento de  Campos e Part\'{\i}culas, Rua Xavier Sigaud
150, Urca, 22290-180,  Rio de Janeiro, RJ, Brasil.}
\item[]{\it \ \ \ e-mail: stephany@cbpfsu1.cat.cbpf.br, arestu@usb.ve,
jlupi@fis.usb.ve}   
\end{description}
\vskip 1cm

{\bf Abstract}
\vskip .5cm

We  study  the interaction  of
non-Abelian topological $BF$ theories defined on  two
dimensional manifolds with point sources carrying non-Abelian charges. We
identify the most general solution for the field equations on simply and
multiply connected two-manifolds. Taking the particular
choice of the so-called extended Poincar\'e group as the gauge group we
discuss how  recent discussions of two dimensional gravity models do fit in
this formalism.

\vskip 2cm 
\hrule 
\bigskip 
\centerline{\bf UNIVERSIDAD SIMON BOLIVAR} 
\vfill
\end{titlepage}

\section{INTRODUCTION}

The adequate description of classical and first quantized relativistic 
objects, like particles and strings, is an essential point of
discussion  in any attempt to set a unified model of  physical
interactions. Our understanding of this issue for relativistic objects has
been continually improved in recent years mainly by the application of BRST
techniques.  In particular,  a  satisfactory relativistic covariant
treatment of isolated scalar \cite{RP} and spinning particles \cite{PS}, and
more recently, superparticles \cite{SP} which allows to discuss most of the
kinematical aspects of such systems have been developed. The interest of
studying  interactions of these systems with classical fields is
manifold.  In a  quantum field theoretical approach, along the lines of
the standard model, new insights on the classical structure
underlying our view of particles interacting with each other through the
interchanging of gauge bosons and gravitons should be of great interest.
They surely would be useful for the  advance in the resolution of the 
formidable technical and conceptual
challenges posed by non perturbative gauge theories and quantum gravity.
In the alternative approach given by Superstring models, a more complete
picture of the classical dynamics of Superparticles and Superstrings
interacting with fields is desirable to elucidate the interaction
mechanism embodied in the vertex operators in order to advance in the
development of a second quantized theory. 

 The case   of particles carrying non-Abelian charge interacting   with
gauge fields is of special interest. For such a system, internal degrees
of freedom in the form of a 
variable $Q(\tau)$ taking values on the algebra are introduced on the
 world line
to take into account the non-Abelian charge.  When the gauge fields are
Yang-Mills fields, the dynamics of the system is derived from the Wong
equations \cite{W}:

\begin{equation}
\ddot{x}^\mu + \Gamma^\mu_{\nu \rho}\dot{x}^\nu \dot{x}^\rho =
-eF^{a\mu}_\nu Q_a(\tau)\dot{x}_\nu (\tau) ,   
\end{equation}

\begin{equation}
{d \over d{\tau}}Q(\tau) + e
\dot{x}^\mu(\tau) \Bigl[A_\mu(x(\tau)),Q(\tau)\Bigr]  = 0 .  
\end{equation}

The first of these equations is the non- Abelian generalization of the 
Lorentz force equation. The second one, as we discuss below in some
detail, may be geometrically interpreted as the parallel transport of the
charge $Q(\tau)$ along the world line.  Due to the non-linearity in
the gauge field, even for flat  space solutions of the Wong equations for the
Yang-Mills system are not easy to find, and discussion of important and
manageable issues in the Abelian case such as radiation and self-interaction is
very difficult to  pursue. 

Recently, systems with alternative dynamics for the gauge fields given
by the  Chern-Simons action in 3 dimensions or the $BF$  
\cite{H} in various dimensions have gained attention.  In both  cases, the pure
gauge theory is topological in the sense that the partition function is metric
independent but the gauge fields can produce definite effects acting on
sources.  In the first case, particles interacting via
Abelian and non-Abelian Chern-Simons fields provide realizations for the novel
possibilities of particle statistics in 3 dimensional space, namely anyonic and
more generally,  Braid statistics \cite{BBJ}. In this paper we will deal with
the second possibility.   

$BF$ models were introduced in Ref.\cite{H} and since then have
attracted wide interest  \cite{BT}.  In particular, their quantization and, the
definition and properties of their observables were the object of much attention
\cite{CGRS}.  In higher dimensions, due to the fact that they are written
in terms of antisymmetric fields or equivalently $p$-forms with $p \ge 2$, $BF$
models couple naturally with extended sources \cite{KR}. In four dimensions
the abelian $BF$ fields induce spin transmutation on stringlike sources
\cite{RG}.

In two dimensions the $BF$ fields
couple to particles. The interaction of the two dimensional abelian $BF$ model
with point sources was discussed in Ref.\cite{LR},  where the general solution
for the field equation was presented. Non-Abelian $BF$ models interacting with
sources  have appeared in the formulation of two  dimensional gravity models as
gauge theories \cite{V}  \cite{CJ92}  \cite{CJ93a}\cite{CJ93b} \cite{GN}
\cite{CGHS}.   These models  written originally in
terms of the geometrical field and an additional dilaton field have been under
study for some time as toy models of Quantum Gravity, and also in connection with String and Superstring models \cite{V}  \cite{CJ92} 
\cite{CJ93a}\cite{CJ93b} \cite{GN} \cite{CGHS}. Their formulation  as gauge
theories \cite{V} \cite{CJ92}  is a step
forward in the study of their quantum properties since it makes available the
well understood machinery of gauge theories and also open space for a better
discussion of the inclusion of non-minimal interactions with matter, a point
which from the physical point of view is of crucial importance when studying 
black hole effects in the theory \cite{CGHS}.

 Our concern in this article
is with the general formulation and solution of the equations
governing the interaction of  $BF$  non-Abelian gauge fields with point
sources carrying non-Abelian charge in 2 dimensional manifolds and their
application to 2 dimensional gravity. In the next section we discuss such
equations starting from the action principle developed in Ref.\cite{BBS}. 
Since for $BF$ theories the gauge connections are flat, we
are able to find the general solution of the equation in simply and
multiply connected surfaces in terms of path ordered operators and step
functions. In the last section we show how our discussion is applied to the
formulation of 2 dimensional gravity  models with sources  as gauge
theories with minimal and non-minimal interactions.
                                                                                                                                                                                                                                                                                                                                                                                                                                                                                                                                                                                                                                                                                                                                                                                                                                                                                                                                                                                                                                                              
\section{NON-ABELIAN BF-THEORIES WITH SOURCES}

Let us consider  a simply connected 2-dimensional surface $M$ described
locally  by coordinates ${\xi^{\mu}}$, (${\mu}=0,1$), and  the principal
fibre  bundle $({\cal P}, M, \pi, G)$ with structure group a Lie group
$G$. Let ${\cal G}$ be the Lie  algebra of $G$, whose generators $T_a$
satisfy the following  relations
\begin{equation}
 [T_a,T_b] = f^c_{ab} T_c  . 
\end{equation}
We introduce an inner product 
\begin{equation}
\label{metric}
<T_a,T_b> = h_{ab}  
\end{equation} 
by means of   an appropriate
non-degenerate,  non-singular matrix $h_{ab}$ invariant in the sense that
\begin{equation}
\label{invariant}
 h_{ab}U^a_cU^b_c=h_{cd}  
\end{equation}
for group elements $U$. The
  indices of the internal space can be raised and  lowered by means of the
 $h_{ab}$. We restrict to the case when 
\begin{equation}
\label{anti}
f_{abc} =f^d_{ab} h_{dc}  
\end{equation} 
is totally antisymmetric. Whenever $G$ is semisimple the inner product is simply
the trace and $h_{ab}$ is the  Cartan-Killing metric 
$h_{ab}=-(1/2)f_{ac}^{d}f_{bd}^c$. In  the case that $G$ is not semisimple, the
Cartan-Killing metric is singular  and the trace is not a suitable inner
product.

	The action that describes the non-Abelian $BF$ theory is 
\cite{H}\cite{BT}
\begin{equation}
S_{BF}=\int_{M^2}<(B \land F)> , 
\end{equation}
where the field $F$ is the curvature 2-form corresponding to 
the connection 1-form $A=A_\mu^a(\xi)T_ad \xi^\mu$ over the
principal  fibre bundle $({\cal P}, M, \pi, G)$, defined by
\begin{equation}
\label{def F}
F= {\cal D}A=dA + e[A,A] , 
\end{equation}
and the field $B=B^a(\xi)T_a$ is a ${\cal G}$-valued
0-form  transforming under the adjoint representation of ${\cal
G}$, and can  be geometrically interpreted as a section of the
vector  bundle $(E, M, \pi_E, {\cal G}, G)$ with typical fibre ${\cal G}$
associated to the  fibre bundle $({\cal P}, M, \pi, G)$ by the adjoint
representation.The operator ${\cal D}=d + e[A,\ ]$ is the gauge covariant 
derivative acting in the adjoint representation \cite{D}, |cite{G}.

	The derivation of Wong equations using a variational principle is
discussed in Ref. \cite{BBS}.  As we mentioned previously the non-Abelian
charge  carried by the particle is described by a ${\cal G}$-valued  variable
$Q(\tau)=Q^a(\tau)T_a$ transforming under the adjoint  representation. In
order to couple the non-Abelian particle  to the gauge field $A$ in a
gauge-invariant manner the following interaction term with support on the
world-line $W$ of the particle is introduced \cite{BBS}, 
\begin{equation}
S_{int}=\int_{W} d\tau <K,g^{-1}(\tau)D_{\tau} g(\tau)>
\end{equation}
with $K=K^aT_a$ being a real constant element of the algebra that, as 
we shall see later on, can be interpreted as an initial
condition and $g(\tau)$  a group element related to the internal variable 
$Q(\tau)$ by
\begin{equation}
\label{charge}
Q(\tau)= g(\tau)Kg^{-1}(\tau) . 
\end{equation}
 The operator $D_\tau$ is the temporal covariant
derivative given by
\begin{equation}
D_\tau= \Bigl[{d \over d\tau} +  e\dot{x}^\mu(\tau)A_\mu(x(\tau)) \Bigr] . 
\end{equation}
   
We are interested in studying the interaction of  a
$BF$ field theory formulated in a 2-dimensional manifold with a set of
non-Abelian charges. We introduce coordinates $x^\mu_i$ and the additional
quantities $g_i$,$Q_i$ and $K_i$ related through (\ref{charge}) and 
incorporate a
coupling term like $S_{int}$ for each particle. The complete action is then
given by 
\begin{equation} 
\label{int}
S= -m\sum_i \int_{W_i} d{\tau}\sqrt{\dot{x}_i^2} + \sum_i\int_{W_i} d{\tau} 
<K_i,g_i^{-1}(\tau)D_\tau g_i(\tau)> + \int_{M^2}<(B \land F)> 
\end{equation}
where $W_i$ are the world-lines of the particles.
The equation of motion  obtained by annihilating the 
variations of $S$ with respect to $B$ is
\begin{equation}
\label{flat}
F=0
\end{equation}
stating that the connection $A$ on the principal bundle $({\cal P}, M, 
\pi, G)$ is flat.

The group  parameters in this formulation are the additional dynamical
variables associated to the non-Abelian charge and have to  be varied
independently.  By varying $g_i$ in the  interaction term, employing
(\ref{metric}), (\ref{invariant}), (\ref{anti}) and (\ref{charge}),
integrating by parts and imposing 
$\delta g_i(\tau_{in})=\delta g_i(\tau_{fin})=0$ we obtain  
\begin{equation}
\label{eq de Q}
{\cal D}Q_i={dQ_i \over d\tau} + 
\dot{x}_i^\mu(\tau)[A_\mu(x_i(\tau)),Q_i(\tau)] =0 
\end{equation}
which is a covariant conservation equation for the non-Abelian charges
$Q_i(\tau)$. It gives the classical configuration for the new dynamical
variables introduced in this case, namely, the group elements.

The  equation obtained varying the coordinates $x_i^\mu(\tau)$ of any of the
trajectories of the particles is the
equation for the geodesics on $M$
\begin{equation}
\ddot{x}_i^\mu + \Gamma^\mu_{\nu \rho}\dot{x}_i^\nu \dot{x}_i^\rho = 0 
\end{equation}
since in this case, due to   the fact that $F=0$, there is no generalized
Lorentz force. Nevertheless as will be show below 
the interaction with the field still forces a  restriction on the allowed set
of geodesics.

Finally, by taking variations with respect to the  gauge field $A^a_\mu$ we
obtain the following equation : 
\begin{equation}
\label{eq de B1}
\epsilon^{\mu \nu}\bigl[\partial_{\nu}B^a +
f_{bc}^{a}A_{\nu}^{b}B^{c}\bigr] + \sum_i\int_{W_i} d\tau
Q_i^a\delta^2(\xi-x_i(\tau))\dot x_i^{\mu}=0 . 
\end{equation}
This can be written also in the form
\begin{equation}
\label{eq de B2}
{\cal D}B(\xi)=(\partial_\mu B +e[A_\mu ,B]) d\xi^\mu = ^*J 
\end{equation}
where $^*$ is the 
Hodge duality operator and $J$
is the current
1-form   associated to the particle 
$$
J=J_\mu ^a T_a d\xi^\mu ,\ \ \ \ \ \ \ \ {^*J}= \epsilon_{\mu \rho}J_\mu ^a T_a
d\xi^\rho ,  
$$ 
\begin{equation}
\label {current}
J^a_{\mu}(\xi)=\sum_i\int_{W_i} d\tau Q_i^a(\tau)\delta^2(\xi-x(\tau))\dot
x_{i \mu}(\tau) . 
\end{equation}

We now turn to the problem of solving 
equations (\ref{eq de Q}), (\ref{eq de B1}) and (\ref{eq de B2}). Since the field $B$ depends
via the  particle current $J$ on the non abelian charges $Q_i$, one 
has to give solutions $Q_i(\tau)$ in order to describe the field $B$.
The equations of motion (\ref{eq de Q}) for 
the non-Abelian charges $Q_i(\tau)$ are parallel-transport equations 
on the vector bundle associated to the principal bundle by 
the adjoint representation. The covariantly conserved non-Abelian 
charges $Q_i(\tau)$ are  then given by the parallel transport 
of  the initial values $Q_i(0)$ lying on the  fibre over $x_i (0)$, along the particle trajectories $W_i$. Therefore in
terms of the path-ordered exponentials we have 
\begin{equation}
\label{sol Q}
Q_i(\tau) =
\left(exp:-\int_{x_i(0)}^{x_i(\tau)}A:\right)Q_i(0)
\left(exp:-\int_{x_i(\tau)}^{x_i(0)}A:\right)
\end{equation} 
from which one can explicitly determine the form of the 
group elements $g_i(\tau)$ appearing in (\ref{charge}). Let us stress again that
the  connection $A$ appearing in (\ref{sol Q}) is flat (cf. eq. \ref{flat}).
For this reason provided that the base  manifold is simply
connected, the ordered exponential  along any closed path is the identity
element. Thus, the  ordered exponentials appearing in (\ref{sol Q}) do not
depend on the  particle trajectory itself but only on its extreme points.

When the base manifold $M$  is multiply connected,
there are  inequivalent solutions \cite{FF}. Let $({\cal P}, M, \pi, G)$ be the principal fibre bundle associated
to $G$, with a  base manifold $M$ which now is a Riemann surface of genus $g
\ge 1$ and $\pi$ denoting the projection map. Let $C$ be a closed  curve on $M$
based on a point $x \in M$ and $p \in {\cal P}$ such that $x=\pi (p)$. 
Parallel transport of $p$ along $C$ results in an  element   
$$ 
exp:-\int_C A:p
$$ 
 belonging to the fibre on $x$ called the
holonomy of  $p$ over the curve $C$ with respect to the connection $A$. 
Varying the closed paths $C$ based on $x=\pi(p)$, the  corresponding
path-ordered integrals are the elements of a  subgroup of $G$, the
holonomy group of the connection  $A$ in $p$. In the
particular case that $A$  is a flat connection, the holonomy group is
isomorphic to  the fundamental homotopy group of $M$ $\pi_1(M)$. Using the
non-triviality of the homotopy group let us  define $Q_i^{mn}(\tau)$  as
\begin{equation}
\label{Qmn} 
Q_i^{mn}(\tau) =
\left(exp:-\int_{C_m}A:\right)Q_i^0(\tau)\left(exp:-\int_{C_n}A:\right),
\end{equation}  
where the ordered exponentials are computed on closed  loops
of the homotopy classes $m$ and $n$ based on $x_i(\tau)$.  Calculating the
dot derivative of $Q_i^{mn}(\tau)$,  it is straightforward to verify  that
$Q_i^{mn}(\tau)$ is also a solution to (\ref{eq de Q})
\begin{equation} 
D_{\tau}Q_i^{mn}(\tau)=0 . 
\end{equation} 
Therefore,  on a multiply connected manifold $M$, the covariant 
conservation equation  $D_{\tau}Q(\tau)=0$ has a family of solutions
\{ $Q_i^{mn}(\tau)$\} each one depending on the homotopy classes $m$, $n$ of
closed  loops $C_m$ and $C_n$ based on $x_i(\tau)$ on which one calculates
the path ordered exponentials.

	We can now construct a solution to the equation 
 (\ref{eq de B2}). 
The field $B$ is a ${\cal G}$-valued 0-form transforming under the 
adjoint representation and can thus  be interpreted as a 
section of the vector bundle $(E, M, \pi_E, G, {\cal G})$ associated to 
the principal bundle $({\cal P} M, \pi, G)$ by the adjoint 
representation. The equation  (\ref{eq de B2}) is non-
homogeneous and  linear in the field $B$. The general 
solution is then the sum of the solution to the homogeneous 
equation $ DB=0$, whose solution is similar in form to the one 
already found for $Q_i(\tau)$, plus a particular solution to the 
non-homogeneous equation. We write at an arbitrary point $P \in M$ 
\begin{equation}
B(P) =
\left(exp:-\int_{P_0}^{P}A:\right)f_W\left(exp:-\int_{P}^{P_0}A:\right) 
\end{equation} 
where $f_W$ is a 0-form depending on the world-lines of the particles $W_i$
that has to be determined, $P_0 \in M$ is a fixed but arbitrary point and the
path ordered integrals are computed over any path connecting $P_0$ and $P$. 
Taking the  exterior derivative of $B$ at the point $P$, using the definition 
of the ordered exponential, and the Leibniz rule one obtains 
\begin{equation}
dB=-AB+BA+\left(exp:-\int_{P_0}^{P}A:\right)df_W
\left(exp:-\int_{P}^{P_0}A:\right) 
\end{equation}     
Using (\ref{eq de Q}) one then has that the 1-form $df_W$ satisfies
\begin{equation}
df_W=\left(exp:-\int_{P}^{P_0}A:\right){^*J}\left(exp:-\int_{P_0}^{P}A:\right) 
\end{equation}
and thus, due to the delta function in the 
current $J$   has support in $\bigcup W_i$. According to 
this, $f_W$ is of constant value at each side of $ W_i$ and is recognized as a
step  function.
 Examining the action of $df_W$ on test 1-forms 
$\varphi$  of compact support one obtains
\begin{equation}
\label{eq df}
\int_{M^2}df_W \land \varphi =\sum_i q_i\int_{W_i} d\tau
\varphi_\mu\left(x(\tau)\right)\dot x_i^\mu(\tau) 
\end{equation}
where $q_i$ are the  quantities
\begin{equation}
\label{df phi} 
q_i:=\left[
\left(exp:-\int_{x_i(\tau)}^{P_0}A:\right)Q_i(\tau)
\left(exp:-\int_{P_0}^{x_i(\tau)}A:\right)\right]
\end{equation}
which, due to the conditions $F=0$ and ${\cal D}_{\tau}Q=0$ are constants.  We note that  if we choose $P_0$
to be $x_i(0)$ then $K_i=Q_i(0)$. In general $K_i$ and $Q_i(0)$ are related by
fixed factors expressed as path ordered exponentials. In obtaining (\ref{df
phi}) we take (\ref{sol Q}) for  simplicity instead of the more
general (\ref{Qmn}).  

The solution to equation (\ref{eq df}) for any test 1-form $\varphi$ do exists
 when  $\bigcup_iW_i$ may be organized in set of paths any of which divide the
manifold and furthermore constitute the boundary of a 2-chain in $M$.   When
this condition is fulfilled the solution is obtained by taking $f_W$ to be
an adequate superposition of Heaviside functions with discontinuty $q_i$
when we cross $W_i$ from right to left. 

In the simplest case of a simply connected unbounded manifold the
physical geodesics are lines coming from the region of the manifold with
minus infinite time coordinate and going to the region with plus infinite time
coordinate. In this case each individual trajectory $W_i$ divide the space in
two sub-manifolds $M_i^+$ and $M_i^-$ and is the boundary of any of them.
The solution for $B$ is then  
\begin{equation}
\label{sol par B}
B_{part}=\left(exp:-\int_{P_0}^{P}A:\right)\sum_iq_i\Theta(M_i^+)
\left(exp:-\int_{P}^{P_0}A:\right) . 
\end{equation}
As explained before a
different choose for $P_0$ amounts to a change in the initial conditions
$Q(0)$.  The  general solution is simply the sum of the particular  solution
(\ref{sol par B}) plus a solution to the homogeneous equation  constructed in a
similar fashion to the one for $Q(\tau)$. 

In the case in which we have only one particle, the  quantity $q\Theta (W)$ in
(\ref{sol par B}) can be expressed alternatively in an  integral form related
to the intersection index of curves  on $M$. Let $C_s$ be an auxiliary curve
transverse to $W$ described  locally by coordinates $z$ with $z(0)=P_0$ and
$z(s)=P_s$. Let us consider now the expression
\begin{equation}
I(C_s,W)=q^{-1}\int _{C_s}\Bigl[
\left(exp:-\int_{P_s}^{P_0}A:\right){^*J}(P_s)\left(exp:-\int_{P_0}^{P_s)}A:
\right)\Bigr] 
\end{equation}
Writing explicitly $^*J(P_s)$ and using (\ref{eq df}) we have that
\begin{equation}
I(C_s,W) = \int_{C_s}ds\int_W \epsilon^{\mu \nu}\dot x^\mu (\tau)z'^\nu (s)
\delta (x(\tau)-z(s))d\tau 
\end{equation}
and  recognize $I(Cs,W)$ as the intersection index between $C_s$ and $W$.
It  is evident that if $C_s$ is closed then $I(Cs,W)=$0. Thus, $I(Cs,W)$ 
is non-zero only if $C_s$ intersects $W$ an odd number of times.  Choosing
$P_0$ at the left of $W$ it is clear that
\begin{equation}
I(C_s,W)=\Theta (W)
\end{equation} 
thus obtaining an integral form for the step function in 
terms of the intersection index. This equation can be generalized for the case of a system of particles by weighting the index with the charges $q_i$ each time we cross $W_i$ from right to left and with $-q_i$ when we cross from left to right.

When the 2-dimensional base manifold $M$ is 
multiply connected the   
trajectories of particles interacting with non-Abelian $BF$ fields are
topologically selected to be homologous to zero as were in the abelian case 
\cite{LR}. To illustrate this point let us consider $M$ to be the manifold
${\bf R}\times S^1$ with $S^1$ identified as the space manifold. The physical
trajectories of a particle are spiral lines coming from $t=-\infty$ and going
to $t=\infty$. The propagation of a single
particle in such manifold is inconsistent with (\ref{eq de B1}), (\ref{eq de B2}). The spiral
line on the cylinder does not divide the manifold and it is not possible to
find $f_W$ to satisfy (\ref{df phi}). The only geodesics on the cylinder that
divide the manifold are the circles $t=constant$ but this are not physical
trajectories. On the other hand if one consider two particles with opposite
non-Abelian charges q moving in parallel trajectories the manifold is divided
in two strips and we may construct the  solution (\ref{sol par B}) with the
support of the Heaviside function on one of the strips.

 In the case that 
the base manifold $M$ is simply connected, given the particle world 
line $W$ and initial conditions $Q_i(0)$ and $B(0)$ the solutions for 
the covariantly conserved non abelian charge $Q_i(\tau)$ and the field 
$B$ are uniquely  determined and given respectively by (\ref{sol Q}) and 
(\ref{sol par B}). Whenever $M$ is multiply connected there will be an 
additional dependence on the homotopy classes of the curves along 
which the ordered exponentials are calculated. In particular, there 
exists a family of solutions {$Q_i^{mn}$} to the equation $D_\tau Q(\tau)=0$
given  by (\ref{Qmn}). From the $Q_i^{mn}$ one constructs the currents
$J^{mn}$  associated to the field equations $DB^{mn}=^*J^{mn}$, with particular 
solutions of the form (\ref{sol par B}).  Let 
us note, that the solution to the homogeneous equation 
$DB_{hom}=0$ has an additional topological information. Just as in the 
case with $Q(\tau)$, one can construct, with the aid of Wilson operators, 
topologically inequivalent solutions
\begin{equation}
\label{B hom} 
B_{hom,pq}(P)=\left(exp:-\int_{P_0}^P
A:\right)_pB(0)\left(exp:-\int_P^{P_0}A:\right)_q  
\end{equation} 
for arbitrary homotopy classes $p$ and $q$. Therefore, for 
solving the homogeneous equation $DB_{hom}=0$ one has to specify 
the initial condition $B(0)$ as well as the homotopy classes of the 
curves associated to the Wilson operators. The difference between two 
solutions $B_{mn}$ and $B'_{mn}$ to the equation $DB_{mn}=^*J_{mn}$ 
is a solution to the homogeneous equation of the form (\ref{B hom}) for 
some $p$, $q$. In this fashion, given a sector associated to homotopy 
classes $m$, $n$ via the current $J_{mn}$, there will be subsectors 
associated to the classes $p$, $q$ of the solutions  of the 
homogeneous equation. Let us note, however, that by varying $p$, 
$q$ it is not possible to jump to another sector associated to a 
current $J_{m'n'}$; this can only be done by modifying 
$Q_{mn} \longrightarrow Q_{m'n'}$. Given a particular solution to 
$DB_{mn}=^*J_{mn}$ one is confined to a sector ($m,n$) even though 
there are several subsectors within associated to the solutions (\ref{B
hom}) to the  homogeneous equation. Other  distinct sectors ($m',n'$) are 
associated  to different field equations via different currents 
$J_{m'n'}$.

\section{APPLICATIONS TO 2-D GRAVITY}

The Einstein theory of gravity in 2 dimensions is trivial since 
the Einstein tensor is in this case identically zero. Various 
alternative models which introduce an additional dilaton field are currently
under study \cite{T} \cite{J}. The string inspired linear gravity
\cite{CGHS},  is specially relevant since
it has a  classical black-hole type solution and a gauge theoretical 
formulation of the model is available \cite{CJ92}. In its original form \cite{CGHS},
the dilaton  gravity model is described by the action 
\begin{equation}
S=\int
d^2x {\sqrt {\overline g}}e^{-2\phi}\left(R+
4{\overline g}^{\mu \nu}\partial_\mu
\phi \partial_\nu \phi - \Lambda \right) 
\end{equation}
where $\Lambda$ is the cosmological constant and $\phi$ is the 
dilaton field. Taking new variables variables $g_{\mu
 \nu}=e^{-2\phi}{\overline g}_{\mu \nu}$ and $\eta=e^{-2\phi}$   \cite{CJ93a}
this can be written in the form
\begin{equation}
\label{action}
S=\int dx^2 {\sqrt g} \left(\eta R -\Lambda \right) . 
\end{equation}
The equation of motion of this action allow classical solutions of the
black-hole type  \cite{CGHS}.

 In what follows we will work with the $Zweibein$  $e^a_{\mu}$, related to
the metric $g_{\mu \nu}$ by
\begin{equation}
g_{\mu \nu}=e^a_\mu e^b_\nu h_{ab} . 
\end{equation}
 As usual, lower case  greek indices refer to the space-time (base) manifold,
take the values  0 and 1, and are raised and lowered by the metric $g_{\mu
\nu}$.  Lower case   indices take the values 0 and 1, refer to the
tangent  space and are raised and lowered with the Minkowski metric 
$h_{ab}=diag(1,-1)$. The antisymmetric tensor is normalized by 
$\epsilon^{01}=1=-\epsilon_{01}$. 
The determinants of the metric and the $Zweibein$ are given 
respectively by $ g  =det g_{\mu \nu}$ and 
${\sqrt -g}=det e^a_\mu = -{1\over 2}e^a_\mu e^b_\nu \epsilon_{ab}$
We shall also need the spin connection $\omega_{\mu}$, and the 
Christoffel symbol $\Gamma^\alpha_{\mu \nu}$ related by
\begin{equation}
\partial_\mu e^a_\nu + \epsilon^a_b \omega_\mu e^b_\nu=\Gamma^\alpha_{\mu
\nu} e^a_\alpha .
\end{equation}
The null-torsion condition
\begin{equation}
\epsilon^{\mu \nu}\left(\partial_\mu e^a_\nu + \epsilon^a_b\omega_\mu
e^b_\nu \right) =0 
\end{equation}
determines the spin connection in terms of the $Zweibein$ and leads to
\begin{equation}
\Gamma^\alpha_{\mu \nu} = {1 \over 2} g^{\alpha \beta}\left(\partial_\mu
g_{\nu \beta} + \partial_\nu g_{\mu \beta} - \partial_\beta g_{\mu
\nu}\right) .  
\end{equation}
Finally, the
scalar curvature $R$ is obtained by
\begin{equation}
\label{Ricci}
\partial_\mu \omega_\nu - \partial_\nu \omega_\mu = - {1 \over 2}{\sqrt
-g} \epsilon_{\mu \nu} R .  
\end{equation}
	
The  gauge  
formulation equivalent to this model discussed in   Ref.\cite{CJ92}  is
based on the extended   Poincar\'e    group,\cite{CJ92} \cite{CJ93a}, whose
Lie algebra reads 
\begin{equation}
[P_a,J]=\epsilon_a^bP_b \  \  \  \  \  [P_a,P_b]= \epsilon_{ab}I \  \  \  \  \  \  \
[J,I]=[P_a,I]=0 
\end{equation}
which  differs from the ordinary  Poincar\'e algebra in that the
translation generators do not commute  due to the presence of the
central element $I$.
 Introducing the notation $T_A=(P_a, J, I)$, $(A=a, 2, 3)$, the algebra 
$[T_A,T_B]=f_{AB}^C T_C$ is 4-dimensional. The Cartan-Killing metric $f_{AD}^C
f_{BC}^D$ is  singular since the group is semisimple; but an invariant,
non-singular,  bilinear form is available with the tensor
\begin{equation}
\label{Cartan}
h_{AB}=\left[ 
\begin{array}{ccc}h_{ab}  & 0 & 0 \\
0  & 0  & -1 \\ 
0  & -1 & 0
\end{array}
\right]   
\end{equation}
which satisfies (\ref{invariant}).
It is employed to raise and lower the algebra indices. One verifies that
$f_{ABC}$ is totally antisymmetric, the only non-zero 
components being the permutations of 
$f_{ab2}=-\epsilon_{ab}$. The invariant product  formed from
(\ref{Cartan}) is thus 
\begin{equation}
\label{product}
\langle W,V\rangle = W^A h_{AB}V^B=W^AV_A=
W^bV_b-W_2V_3-W_3V_2=W^bV_b-W^3V^2-W^2V^3 
\end{equation}
In the dilaton gravity model (\ref{action}) the dynamical variables are  the
multiplier $\eta$ and the metric $g_{\mu \nu}$ (or equivalently,  the
dilaton $\phi$ and the metric). In the equivalent extended Poincar\'e   
gauge formulation one introduces a connection 1-form  $A$ on a principal
bundle with the extended  Poincar\'e  group as  structure group, whose
expression on the base manifold is 
\begin{equation}
A=A_\mu^Ad\xi^\mu T_A=e_\mu^a d\xi_\mu P_a +\omega_\mu d\xi^\mu J + a_\mu
d\xi^\mu I . 
\end{equation}
The components of this gauge potential are the dynamical variables of the 
theory: the $Zweibein$ $e^a_{\mu}$, the spin connection 
${\omega_{\mu}}$, and the potential $a_{\mu}$.
Parameterizing the group element $g$ in the form
\begin{equation}
g=e^{\theta_aP^a}e^{\alpha J}e^{\beta I}
\end{equation}
the transformation laws for the components are obtained for the usual rule
for the connection.
We have
$$
\omega_\mu \rightarrow \omega_\mu + \partial_\mu \alpha
$$
\begin{equation}
\label{diff}
e^a_\mu \rightarrow \left(L^{-1}\right)^a_b\left(e^b_a+\epsilon^b_c
\theta^c \omega_\mu + \partial_\mu \theta^b \right)
\end{equation}
$$
a_\mu \rightarrow a_\mu - \theta^a \epsilon_{ab}e^b_\mu -{1 \over 2}
\theta^a \theta_a \omega_\mu + \partial_\mu \beta +{1 \over 2} \partial_mu
\theta^a\epsilon_{ab} \theta^b
$$
where $L^a_b$ is the Lorentz transformation of rapidity $\alpha$
\begin{equation}
L^a_b= \delta^a_b cosh \alpha + \epsilon^a_b sinh \alpha
\end{equation} 

 From $A$ one 
constructs the curvature 2-form $F$ 
\begin{equation}
F=dA+A\land A =F^AT_A = (de^a+\epsilon^a_b\omega e^b)+d\omega J+(da+{1
\over 2}e^a\epsilon_{ab}e^b)I = f^aP_a + d\omega +\nu I .  
\end{equation}
  The components along the translations and rotations are the torsion  and
the scalar curvature respectively. The two terms in the  component along
$I$ are the field strength associated to the  potential $A$ and the
volume element expressed in terms of the  $Zweibein$. By incorporating a
Lie algebra-valued Lagrange  multiplier $\eta^A=(\eta^a, \eta^2, \eta^3)$
and using the inner  product (\ref{product})  one constructs the $BF$-type
action 
\begin{equation}
\label{action BF}
S_{\eta F}=\int<\eta,F> .  
\end{equation}
This action has been shown to be classically equivalent to 
(\ref{action}). 

A consistent interaction of this system with particles may be obtained
directly from (\ref{int}) and our discussion of the previous section 
(identifying the $B$ field with the multiplier $\eta$). For a single particle
we have
\begin{equation} 
\label{bf}
S= -m \int_{W} d{\tau}\sqrt{\dot{x}^2} + \int_{W} d{\tau} 
<K,g^{-1}(\tau)D_\tau g(\tau)> + \int_{M^2}<(\eta , F)> . 
\end{equation}

 We have to take care of the fact
that the kinetic term of the  the action for the  particle,
explicitly includes the metric $g_{\mu \nu}$  evaluated in $x(\tau)$ and
therefore, since the $Zweibein$ is a  dynamical variable as a component of the
connection $A$, it is  necessary to consider the extra contribution of this
term while  performing variations with respect to $A$. This is done by
observing that the corresponding field equation may be written in the form:
\begin{equation}
\label{eq de eta}
D_\mu \eta = \epsilon_{\mu \nu}\int_W d\tau \widetilde Q
\delta^2(\xi-x(\tau))\dot x^\nu
\end{equation}
where 
\begin{equation}
\label{hatQ}
\widetilde Q=Q-{m\over \sqrt{\dot{x}^2}}\dot x^\mu e^a_\mu P_a .
\end{equation}
The other field equations are the zero curvature condition (\ref{flat}) and
the covariant conservation of $Q(\tau)$. The important point is that then
$\widetilde Q$  also satisfies
\begin{equation}
\label{eq de hatQ}
D_\tau \widetilde Q =0
\end{equation}
 since the extra term in (\ref{hatQ})
is also conserved as can be verified by direct computation. So the
solution to (\ref{eq de eta}) may be found directly from (\ref{sol par B})
and the expression for (\ref{hatQ}) is given by (\ref{sol Q}) 
	
 Let us consider the solution to the equation 
(\ref{eq de hatQ}) with initial condition $Q(0)=Q^a_0P_a+Q^3_0I$ 
\begin{equation}
\label{sol hatQ}
\widetilde Q(\tau) =
\left(exp:-\int_{x(0)}^{x(t)}A:\right)(Q^a_0P_a+Q^3_0I)
\left(exp:-\int_{x(t)}^{x(0)}A:\right)
\end{equation} 
and let us calculate explicitly the components $\widetilde Q^A$ (This is done
for later use when we will  to show the equivalence with the formulation
of  Cangemi and Jackiw). Since the central element $I$ commutes with all  the
generators our labour is simplified. We are then interested in computing
\begin{equation}
(1-A\Delta)_n\cdots(1-A\Delta)_1P_a(1-A\Delta)_1\cdots (1-A\Delta)_n
\end{equation} 
using the definition of the path-ordered exponential as 
the limit $\Delta \rightarrow 0$ of the product of factors $(1-
A\Delta)_i$, $\Delta=\Delta p_i$ with $p_i$ being  points on the path 
$W$ and the subindex $i$ meaning that the expression between 
parentheses is evaluated in $p_i$. We fix $x(0)=p_1$ and 
$x(\tau)=p_n$. Noting that
\begin{equation}
(1-A\Delta)_1P_a(1-A\Delta)_1 = \left(\delta^a_b + \epsilon_a^b
\omega\Delta\right)_1P_b +\epsilon_{ab}(e^b\Delta)_1I + O(\Delta^2)
\end{equation}
 we have 
\begin{equation}
\prod^{n-1}_{i=1}\left(\delta^{a^{i+1}}_{a_i}+\epsilon^{a_{i+1}}_{a_i}
\omega\Delta\right)_iQ_0^{a_{1}}P_{a_{n}}+
\sum_{i=1}^n\Delta\epsilon_{a_{i}b}(e^b)_i
\left[\prod^i_{j=1}\left(\delta^{a_{j}}_{a_{j-1}}+\epsilon^{a_{j}}_{a_{j-1}}
\omega\Delta\right)_{j-1}Q_0^{a_{1}}\right]I
\end{equation}

Taking  $\Delta \rightarrow 0$ it becomes
\begin{equation}
\label{sol Qa}
\widetilde Q^a(\tau)P_a=exp:-\int_{x(0)}^{x(\tau)}\epsilon_b^a\omega:Q_0^bP_a
\end{equation}
\begin{equation}
\label{sol Q3}
\widetilde Q^3(\tau)I=\int_0^\tau e^a(s)p_a(w(s))I+Q_0^3 I
\end{equation}

An alternative method for switching on the interaction was originally
proposed by Cangemi and Jackiw \cite{CJ93a}\cite{CJ93b} introducing the
degrees of freedom associated to the non-Abelian charge in a different way.
They construct their formulation in terms of a set of  tangent space
coordinates and momenta $q^a$ and $p_a$. That such  formulation may be 
equivalent to construction  presented
in the previous section is suggested by the fact discussed in \cite{BBS} that
the ${\cal G}$-valued variables $Q$ are related by a canonical transformation
to the conjugate momenta of the group elements $g$ used in the lagrangian
formulation. Actually we will show presently that (\ref{sol Qa}) and (\ref{sol
Q3}) are solutions to the equations of the system discussed by Cangemi and
Jackiw
\cite{CJ93a},\cite{CJ93b} but we stress that they have the advantage that can be
easily generalized to global solutions in multiply connected manifolds. 

In its most compact
\cite{CJ94} form, the particle action of Cangemi and Jackiw is written in the
form

\begin {equation}
\label{action CJ}
S_{p}=\int d\tau \left[p_a(D_\tau q)^a-{1\over 2} N(p^2+m^2) \right]
\end{equation}
where $(D_\tau q)^a=\dot{q}^a+\epsilon^a_b(q^b\omega_\mu -e^b_\mu)\dot
x^\mu$,  $x^\mu(\tau)$ are the trajectory coordinates on the manifold and
$q^a$ are the Poincar\'e parameters identified as mentioned with the
coordinates of tangent space. This action is invariant under the gauge
transformation (\ref{diff}) if $q^a$ and $p_a$ transform like
\begin{equation}
q^a \rightarrow (\lambda^{-1})^a_bq^b + \rho^a, \ \ p_a \rightarrow
\lambda^b_ap_b
\end{equation}
with $\lambda$ and $\rho$ evaluated on the trajectory. 

The interacting system defined by (\ref{action BF}) and (\ref{action CJ})
\begin{equation}
S_{CJ}=S_{\eta F}+ S_{p}
\end{equation}
leads to the flat curvature condition $F=0$ and the equations of motion
\begin{equation}
\dot{p}_a+\epsilon_a^bp_b \omega_\mu \dot{x}I \mu =0
\end{equation}
\begin{equation}
\label{eq q}
(D_\tau q)^a =Np^a
\end{equation}
and
\begin{equation}
\partial_\mu \eta + [A_\mu,\eta]=\epsilon_{\mu \nu}J^\nu .
\end{equation} 
The matter current $J^\mu$ may be written in the form 
\begin{equation}
J^\mu(\xi) = \int d\tau j \delta^2(\xi-x(\tau)\dot{x}^\mu(\tau)
\end {equation}
with the components in $j=j^aP_a+j^2J+j^3I$ such that
\begin{equation}
\label{comp j}
j=-\epsilon^{ab}p_bP_a-\epsilon_a^bq^ap_bI .
\end{equation}
Then then using (\ref{eq q}) we have
\begin{equation}
{dj^a \over d\tau} + \epsilon^a_b \omega_\mu \dot{x}^\mu j^b=0
\end{equation}
\begin{equation}
{dj^3 \over d\tau} - e^a_\mu \dot{x}^\mu p_a=0 .
\end{equation}

The solutions of this equations  are given by 
(\ref{sol Qa}), (\ref{sol Q3}) if we make 
the following identifications  
\begin{equation}
\widetilde Q^a(\tau)=j^a(\tau)=-\epsilon^{ab}p_b ,
\end{equation} 
\begin{equation}
\widetilde Q^3(\tau)=j^3(\tau)=-\epsilon_a^bq^ap_b .
\end{equation}

The treatment described above can also be applied to the more general
situation studied in \cite{CJ93b} where non-minimal gravitational
interactions are considered. Let us consider the gauge invariant action 
\cite{CJ93b}
\begin{equation}
\label{non minimal}
S_t=\int \eta_A F^A + \int d\tau \left[ p_a(D_\tau q)^a - {1 \over 2}
N(p^2 + m^2) + q_a A^A_\mu \dot{x}^\mu - {1 \over 2}q^a \epsilon_{ab}q^b
\right] \end{equation}  
where now $q^{3}=(1/2)q^{a}q_{a}+{\cal A}$
(${\cal A}$ is a constant). The extra terms in (\ref{non minimal}) introduce a
forcing term in the geodesic equation arising solely from gravitational
variables. After choosing the "physical gauge" $q^a=0$, performing
variations with respect to $x^\mu$ leads to the geodesic equation, modified
by a gravitational force \cite{CJ93b} 
\begin{equation} 
{d \over d\tau}{1 \over N}\dot{x}^\mu + {1 \over N}\Gamma^\mu _{\alpha
\beta} \dot{x}^\alpha \dot{x}^\beta +({1 \over 2}{\cal A}R +1)g^{\mu
\alpha}{\sqrt g}\epsilon_{\alpha \beta} \dot{x}^\beta =0 .
 \end{equation} 
The term including ${\cal A}R$ is non-minimal and vanishes in flat-spacetime.
The second term resembles the interaction with an external electromagnetic
field in two dimensions. In our framework, this additional interaction can be
incorporated by adding to the action (\ref{int}) the gauge-invariant term
\begin{equation}
-\int d^2\xi \int d\tau \delta^2(\xi-x(\tau))<J+{\cal A}I,A_\nu> .
\end{equation}
Using (\ref{Ricci}) one can verify that the interaction term
above originates, upon variation with respect to $x^\mu$, the forced geodesic
equation (\ref{non minimal}). The equation of motion for the multiplier $\eta$
is again (\ref{eq de eta}) but now  
\begin{equation} 
\widetilde{Q}=Q-{\dot{x}^\mu \over N}e^a_\mu P_a - J - {\cal A}I
\end{equation} 
and once more, after some algebra, one gets $D_{\tau}\widetilde{Q}=0$. In order to obtain
$\widetilde{Q}(\tau)$ we proceed as above with the initial condition
$\widetilde{Q}(0)=Q^{a}_{¯}P_{a}+Q^{2}_{¯}J+Q^{3}_{¯}I$. The
components of $\widetilde{Q}(\tau)$ proportional to $P_a$ and $J$ are respectively
$$
\widetilde{Q}^a(\tau)=\left(exp:-\int_0^t\epsilon
\omega:\right)_b^aQ_0^bP_a-\left(exp:-\int_0^t \epsilon\omega:\int_0^tds
exp:\int_0^s\epsilon\omega:\epsilon e(s)\right)^a Q_0^2P_a , 
$$
\begin{equation}
\label{ur}
 \widetilde{Q}^2(\tau)=Q_0^2J .
\end{equation}

Let us  compare again with the formulation of Cangemi and
Jackiw. The equations of motion obtained from the action (\ref{non minimal}) 
besides $F=0$ are 
$$
D_\tau q= N(p+<p,q>I),
$$
$$
D_\tau p+<D_\tau p,q>I=[D_\tau q,q]
$$
\begin{equation}
\label{eqsnm} 
D_\mu \eta = \epsilon_{\mu \nu} J^\nu
\end{equation} 
where the current $J^{\nu}$ is
given by 
\begin{equation} 
J^\nu(\xi) = \int d\tau([p(\tau),q(\tau)]-q(\tau)){\dot
x}(\tau)\delta^2(\xi-x(\tau))   
\end{equation} 
From the equations of
motion (\ref{eqsnm}) above, we note that $D_{\tau}([p,q]-q)=0$, so that it is
possible to identify $[p,q]-q$ with $ \widetilde{Q}(\tau)$. In fact, defining
$[p,q]-q=h^{A}T_{A}$, one gets that the components $h^{a}$ and $h^2$ satisfy 
$$
{\dot h}^2=0
$$
\begin{equation} 
{\dot h}^a = -({\dot q}^a +\epsilon^{ab}{\dot p}_b)=(-\epsilon_b^a\omega h^b
+\epsilon_b^a e^b) 
\end{equation} 
The solutions to these equations are precisely the
expressions (\ref{ur}) for the components $ \widetilde{Q}^a$ and $ \widetilde{Q}^2$ obtained when solving
the covariant conservation equation for $ \widetilde{Q}(\tau)$ with the aid of the
path-ordered exponentials. In this fashion, we are allowed to identify
$ \widetilde{Q}^{a}(\tau)=([p,q]-q)^{a}=h^{a}$ and $ \widetilde{Q}^{2}_{¯}=([p,q]-q)^{2}=h^{2}=-1$.
The expression for $ \widetilde{Q}^{3}(\tau)$ is calculated by taking into account the
terms proportional to $I$, and the identification with $h^3$ is obtained
after following the procedure shown above.

\section{DISCUSSION}

The method
proposed in Ref.\cite{BBS} in the context of Yang-Mills theory provides a
general rule for the coupling of non-Abelian point sources with gauge fields.
Following this approach we constructed the action describing the interaction of non-Abelian point sources with a set of $BF$ fields. For this system the condition $F=0$ determines that the particles suffer no
Lorentz like forces and this allows the equations of motion to be integrated
explicitly. The solutions for the $B$ fields may then be written in terms of 
step functions depending only on the trajectories of the particles which are
geodesics. Using the equivalence of the two dimensional gravity models
with corresponding $BF$ models, this results lead to a new formulation of the interaction of two dimensional gravity with sources in terms of
the geometrical variables which is equivalent to  the results of Ref.\cite{CJ92}
\cite{CJ93b} .

\section{ACKNOWLEDGMENTS}

We thank Prof. R.Jackiw for pointing out to us the application to 2D gravity. 
J.S. thanks the Centro Latinoamericano de F\'{\i}sica  (CLAF) and the Conselho Nacional de Desenvolvimento Cient\'{i}fico e Tecnol\'ogico (CNPq) for financial support.
\vskip .5cm

{\it Note added in proof}: After the completion of this work we became aware of Ref.\cite{CD} where Lie valued variables are also considered for the description of matter in 2D gravity.

\end{document}